\documentclass[11pt]{article}
\usepackage{vietnam,epsfig,psfrag}

\def\be{\begin{equation}}
\def\ee{\end{equation}}
\def\bea{\begin{eqnarray}}
\def\eea{\end{eqnarray}}

\psfrag{lij}{$\lambda_{ij}$}
\psfrag{th1}{$\theta_1$}
\psfrag{th2}{$\theta_2$}
\psfrag{mN}{$m_N$}
\psfrag{mol}{$|\lambda_{ij}|$}
\psfrag{a}{$a$}
\psfrag{l}{$\lambda_{23}$}
\psfrag{Htm}{$BR(H_0 \to \tau \bar \mu)$}
\psfrag{htm}{$BR(h_0 \to \tau \bar \mu)$}
\psfrag{Atm}{$BR(A_0 \to \tau \bar \mu)$}
\psfrag{meg}{$BR(\mu \to e \gamma)$}
\psfrag{tmg}{$BR(\tau \to \mu \gamma)$}
\psfrag{ttm}{$BR(H_x \to \tau \bar \mu)$}
\psfrag{mth1}{$|\theta_1|$}
\psfrag{mth2}{$|\theta_2|$}
\psfrag{mth3}{$|\theta_3|$}
\psfrag{M0}{$M_0$}
\psfrag{M12}{$M_{1/2}$}
\psfrag{Msu}{$M_{SUSY}$}
\psfrag{lk}{$l_k$}
\psfrag{lm}{$l_m$}
\psfrag{blk}{$\bar {l}_k$}
\psfrag{blm}{$\bar {l}_m$}
\psfrag{s}{$a$}
\psfrag{la}{$\tilde {l}_{\alpha}$}
\psfrag{lb}{$\tilde {l}_{\beta}$}
\psfrag{Hsll}{$BR(H_{SM} \to l_k \bar{l}_m)$}
\psfrag{LBR_SM}{$\log \left(BR(H_{SM} \to l_k \bar{l}_m)\right)$}
\psfrag{LmN}{$\log(m_N)$}
\psfrag{argth1}{$Arg(\theta_1)$}
\psfrag{argth2}{$Arg(\theta_2)$}
\psfrag{argth3}{$Arg(\theta_3)$}
\psfrag{M0M12}{$M_0=M_{1/2}$}
\psfrag{HAtm}{$BR(H_0,A_0 \to \tau \bar \mu)$}
\begin{document}
%\vspace*{1cm}
\title{Lepton flavor violating Higgs boson decays\\ in the MSSM-seesaw}

\author{E. ARGANDA$^{(a)}$, A. M. CURIEL$^{(a)}$, M. J.
HERRERO$^{(a)(b)}$ and D. TEMES$^{(c)}$}

\address{(a) Dpto. de F\'{\i}sica Te\'orica, Universidad Aut\'onoma de Madrid, Spain.\\
(b) Talk given at the Vth Rencontres du Vietnam, Hanoi, August 5-11 2004.\\
(c) Laboratoire de Physique Th\'eorique, LAPTH, France.}
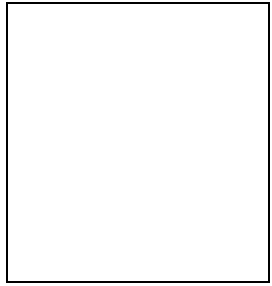
\maketitle\abstracts{Lepton flavor violating Higgs boson decays (LFVHD) are studied in the 
context of the Minimal
Supersymmetric Standard Model (MSSM) enlarged with three right handed neutrinos 
 and their
supersymmetric partners, and with the neutrino masses being generated by the seesaw mechanism. 
We compute the partial widths for these decays to one-loop order
and analyze numerically the corresponding branching ratios
in terms of the MSSM and seesaw parameters. 
We analyze in parallel the lepton flavor
changing $\l_j\to l_i \gamma$ decays and explore the maximum predicted rates for 
LFVHD, mainly for $H^0,A^0 \to \tau {\bar \mu}$ decays, by requiring compatibility with neutrino
and $BR(\l_j\to l_i \gamma)$ data. We find LFVHD ratios of up to $10^{-5}$ in some regions of the
MSSM-seesaw parameter space.} 
\section{Introduction}
The observed neutrino masses do require a theoretical framework beyond the Standard Model of
Particle Physics with just three massless left-handed neutrinos.
Within the MSSM-seesaw context, which will be adopted here, the MSSM particle content is enlarged
by three right handed neutrinos plus their corresponding supersymmetric (SUSY) partners, and the 
neutrino masses are generated by the seesaw mechanism. Three of the six resulting Majorana
neutrinos have light masses, $m_{\nu_i}, i=1,2,3$, and the other three have heavy  masses, 
$m_{N_i}, i=1,2,3$. These physical masses are related to the Dirac mass matrix 
$m_D$, the right-handed neutrino mass matrix $m_M$, and the unitary matrix $U_{MNS}$ by 
$diag(m_{\nu_1}, m_{\nu_2}, m_{\nu_3})\simeq U_{MNS}^T(-m_Dm_{M}^{-1}m_D^T)U_{MNS}$ and 
$diag(m_{N_1}, m_{N_2}, m_{N_3})\simeq m_M$, respectively. Here we have chosen an electroweak 
eigenstate basis where $m_M$ and the charged lepton mass matrix are flavor diagonal, and we
have
assumed that all elements in $m_D=Y_{\nu}<H_2>$, where $Y_{\nu}$ is the neutrino Yukawa
coupling matrix and $<H_2>=v \sin \beta$ ($v=174$ GeV), are much smaller than those of $m_M$. The two
previous relations can be rewritten together in a more convenient form for the work presented
here as, 
$m_D^T =i m_N^{diag \, 1/2} R m_{\nu}^{diag \, 1/2} U_{MNS}^+$, where $R$ is a general complex
and orthogonal $3 \times 3$ matrix, which will be parameterized by three complex angles
$\theta_i$, $i=1,2,3$. 

One of the most interesting  features of the MSSM-seesaw model is the associated rich
phenomenology due to the occurrence of lepton flavor violating (LFV) processes. 
Whereas in the 
standard (non-SUSY) seesaw models the ratios of LFV processes are small due to the smallness of
the light neutrino masses, in the SUSY-seesaw models these can be large due to an important
additional source of lepton flavor mixing in the soft-SUSY-breaking terms. Even in the
scenarios with universal soft-SUSY-breaking parameters at the large
energy scale associated to the SUSY breaking $M_X$, the running from this scale down to $m_M$ 
induces, via the neutrino Yukawa couplings, large lepton flavor mixing in the slepton 
soft masses, 
and provides the so-called slepton-lepton misalignment, which in turn
generates 
non-diagonal lepton flavor interactions. 
These interactions can induce sizable ratios in several LFV processes with SM charged 
leptons in the 
external legs, which are actually being tested experimentally with high precision and therefore provide 
a very interesting window to look for indirect SUSY signals. They can also induce important
contributions to other LFV processes that could be meassured in the next generation 
colliders, as it is the case of the MSSM Higgs boson decays into $\tau {\bar \mu}$,     
$\tau {\bar e}$ and $\mu {\bar e}$ which are the subject of our interest.

Here we compute the partial widths for these lepton flavor violating Higgs boson decays (LFVHD)
to one-loop order
and analyze numerically the corresponding branching ratios
in terms of the MSSM and seesaw parameters, namely, $M_0$,
$M_{1/2}$, $\tan\beta$, $m_{N_i}$ and $R$.  
We analyze in parallel the lepton flavor
changing $\l_j\to l_i \gamma$ ($i\neq j$) decays and explore the maximum predicted rates for 
LFVHD, mainly for $H^0,A^0 \to \tau {\bar \mu}$ decays, by requiring compatibility with 
$BR(\l_j\to l_i \gamma)$ data. For these we use the present
experimental upper bounds given by~\cite{todas}  
$|BR(\mu \to e \gamma)|<1.2 \times 10^{-11}$, 
$|BR(\tau \to \mu \gamma)|<3.1 \times 10^{-7}$ and 
$|BR(\tau \to e \gamma)|<2.7 \times 10^{-6}$.

For the numerical analysis we choose, $M_X=2\times 10^{16}$ GeV and $A_0=0$. The $U_{MNS}$ 
matrix
elements and the $m_{\nu_i}$ are fixed
to the most favored values by neutrino data~\cite{review} with  
$\sqrt{\Delta m_{sol}^2}=0.008$ eV, $\sqrt{\Delta m_{atm}^2}=0.05$ eV, 
$\theta_{12}=\theta_{sol}=30^o$, $\theta_{23}=\theta_{atm}=45^o$, 
$\theta_{13}=0^o$ and $\delta = \alpha= \beta =0$. We consider two plaussible scenarios, 
one with
quasi-degenerate light and degenerate heavy neutrinos and  with    
$m_{\nu_1}=0.2 \, eV \, , m_{\nu_2}=m_{\nu_1}+\frac{\Delta m_{sol}^2}{2 m_{\nu_1}} 
\, , m_{\nu_3}=m_{\nu_1}+\frac{\Delta m_{atm}^2}{2 m_{\nu_1}}$ and 
$m_{N_1} = m_{N_2}= m_{N_3}= m_N$; and the other one with hierarchical light and hierarchical
heavy neutrinos, and with 
$m_{\nu_1} \simeq 0 \, eV \, , m_{\nu_2}= \sqrt{\Delta m_{sol}^2} \, , 
m_{\nu_3}=\sqrt{\Delta m_{atm}^2}$ and
$m_{N_1} \leq  m_{N_2} < m_{N_3}$. 

This is a reduced version of our more complete work~\cite{nosotros} to which we address the
reader for more details.
\section{Numerical results and conclusions}
\begin{figure}
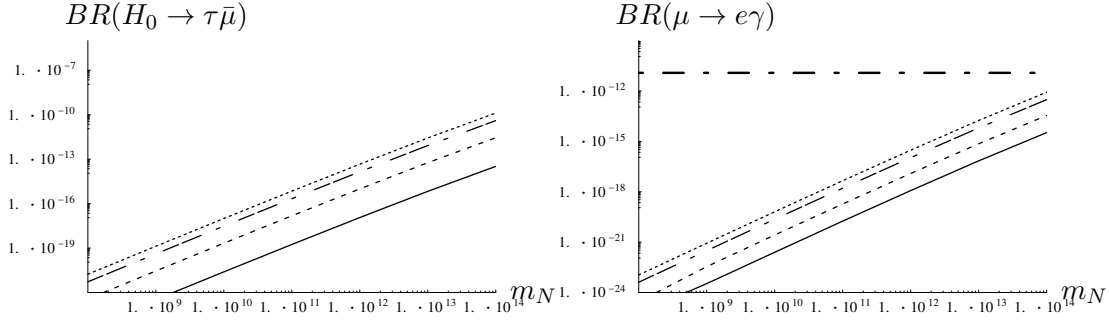

\psfig{figure=BR_H0_mN_log.nb.epsi,height=1.75in}
\psfig{figure=mN_mas_puntos_plots.nb.epsi,height=1.75in}
%\psfig{figure=BR_A0_mN_log.nb.epsi,height=1.75in}
%\hspace{1.75cm}
%\psfig{figure=BR_h0_mN_log.nb.epsi,height=1.75in}
\caption{Dependence of LFV ratios with $m_N$ (GeV) 
for degenerate heavy neutrinos and real R and for several values of 
$\tan\beta$. {\bf (1a)} $BR(H_0 \to \tau \bar \mu)$. 
 {\bf (1b)} $BR(\mu \to e \gamma)$. The horizontal line 
is the experimental upper bound. In both plots,
the solid, dashed, dashed-dotted and  dotted lines are the preditions for  
$\tan \beta=3,10,30$ and $50$, respectively,   
and
$M_0=400 $ GeV, $M_{1/2}=300 $ GeV.
\label{fig:1}}
\end{figure}

\begin{figure}
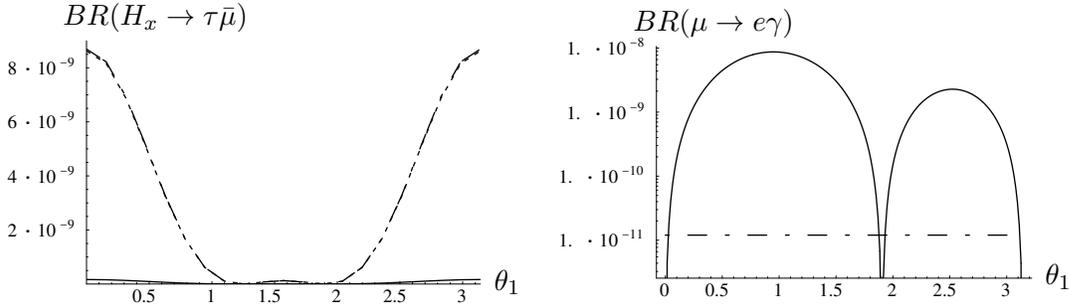

\psfig{figure=BR_angulo1_real.tanb50.nb.epsi,height=1.75in}
\psfig{figure=hisano_angulo1_real.tanb50.nb.epsi,height=1.75in}
%\psfig{figure=BR_angulo2_real.tanb50.nb.epsi,height=1.75in}
%\hspace{1.75cm}
%\psfig{figure=hisano_angulo2_real.tanb50.nb.epsi,height=1.75in}
\caption{LFV ratios for hierarchical heavy neutrinos and real $R$ with $\theta_1 \neq 0$,  
$\theta_2=\theta_3=0$ and $(m_{N_1},m_{N_2},m_{N_3})=(10^8, 2 \times 10^8, 10^{14})$ GeV. 
Here, $\tan \beta = 50$, $M_0=400 $ GeV and $M_{1/2}=300 $ GeV.
{\bf (2a)} $BR(H_x \to \tau \bar \mu)$. 
Solid, dashed and dashed-dotted lines (the two later undistinguishible here and in all plots) correspond to 
$H_x=(h_0, H_0, A_0)$ respectively. 
{\bf (2b)} $BR(\mu \to e \gamma)$. The horizontal line is 
the experimental upper bound.
\label{fig:2}}
\end{figure}

\begin{figure}
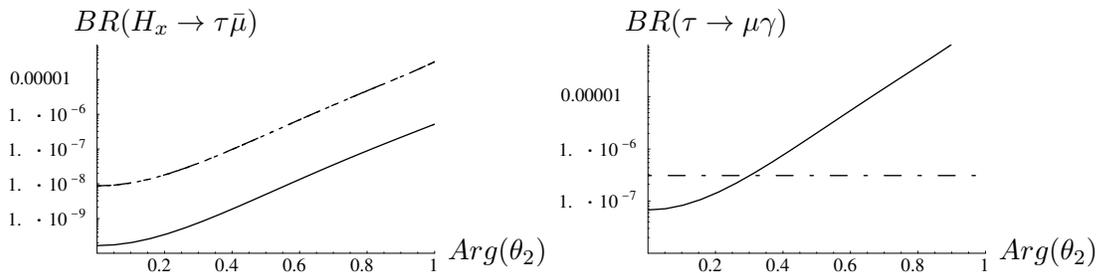

\psfig{figure=BR_angulo2_complex.logscale.arg.nb.epsi,height=1.75in}
\psfig{figure=hisano.angulo2_complex.arg.nb.epsi,height=1.75in}
%\psfig{figure=BR_angulo2_complex.logscale.mod.nb.epsi,height=1.75in}
%\hspace{1.75cm}
%\psfig{figure=hisano.angulo2_complex.mod.nb.epsi,height=1.75in}
\caption{LFV ratios for hierarchical heavy neutrinos and complex $R$ with $\theta_2 \neq 0$,  
$\theta_1=\theta_3=0$, and $(m_{N_1},m_{N_2},m_{N_3})=(10^8, 2 \times 10^8, 10^{14})$ GeV. 
Here, $\tan \beta = 50$, $M_0=400 $ GeV, and $M_{1/2}=300 $ GeV.
{\bf (3a)} Dependence of $BR(H_x \to \tau \bar \mu)$ with 
$Arg(\theta_2)$ for $|\theta_2|=\pi$. Solid, dashed and dashed-dotted 
lines are for $H_x=(h_0, H_0, A_0)$ 
respectively.
{\bf (3b)} Same as (3a) but for $BR(\tau \to \mu \gamma)$.  
The horizontal line is the experimental upper bound.
\label{fig:3}}
\end{figure}

\begin{figure}[t]
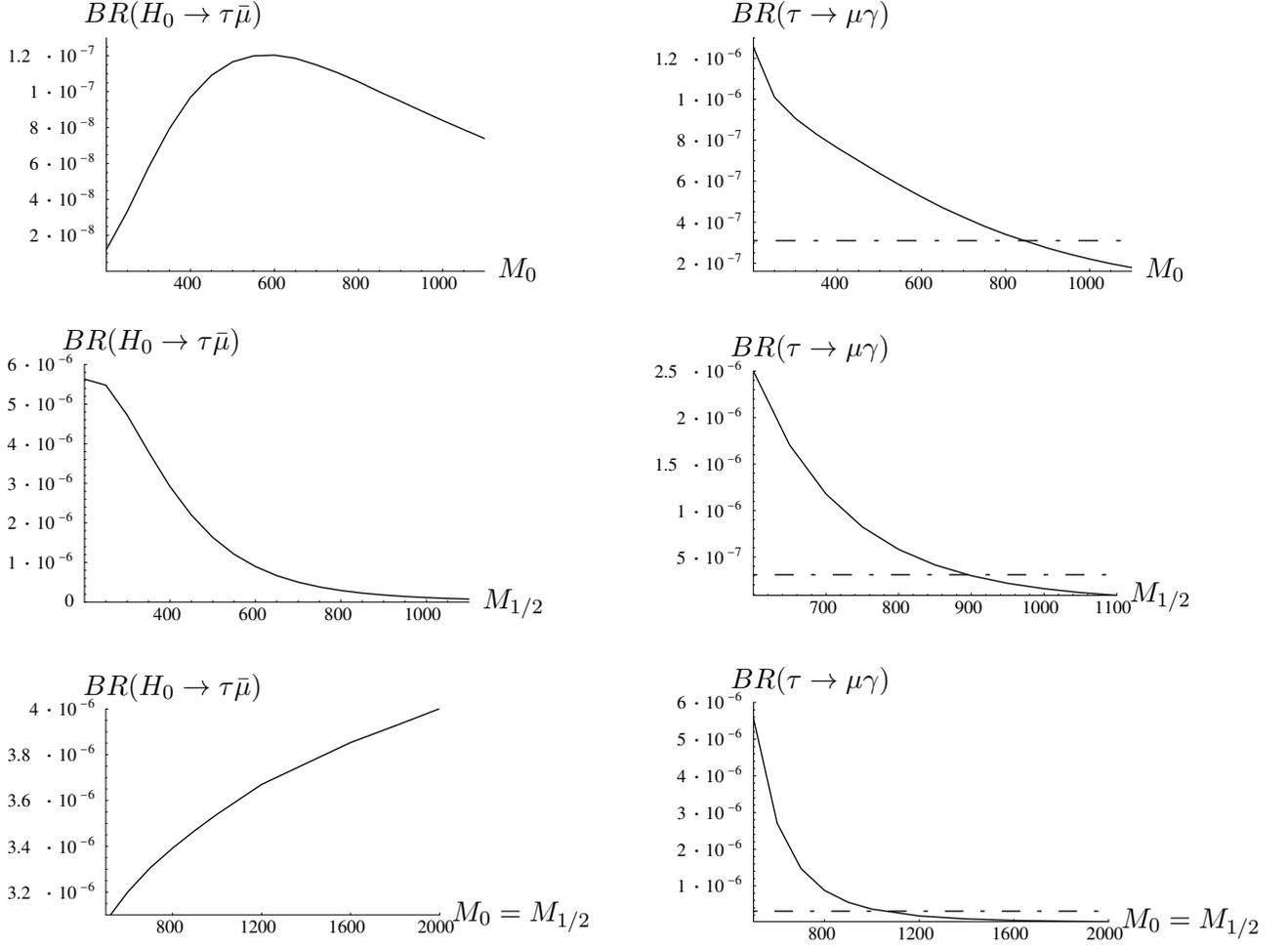

\psfig{figure=BR_M0.nb.epsi,height=1.75in}
\psfig{figure=hisano.M0.nb.epsi,height=1.75in}
\psfig{figure=BR_M12.nb.epsi,height=1.75in}
\psfig{figure=hisano.M1_2.nb.epsi,height=1.75in}
\psfig{figure=BR_M0.igualM1_2.nb.epsi,height=1.75in}
\hspace{1.75cm}
\psfig{figure=hisano.M0.igualM1_2.nb.epsi,height=1.75in}
\caption{Dependence with $M_0$ (GeV) and $M_{1/2}$ (GeV)
for $(m_{N_1},m_{N_2},m_{N_3})=(10^{8},2 \times 10^{8}, 10^{14})$ GeV, $\theta_1=\theta_3=0$ and 
$\tan \beta = 50$.
{\bf (4a)} $BR(H_0 \to \tau \bar \mu)$ versus $M_0(GeV)$ 
for 
$M_{1/2}=300$ GeV and $\theta_2 =\pi e^{0.4i}$. 
{\bf (4b)} Same as (4a) but for $BR(\tau \to \mu \gamma)$. 
{\bf (4c)} $BR(H_0 \to \tau \bar \mu)$ versus
$M_{1/2}(GeV)$ for
$M_0=400$ GeV and $\theta_2 =\pi e^{0.8i}$. 
{\bf (4d)} Same as (4c) but for $BR(\tau \to \mu \gamma)$. 
{\bf (4e)} $BR(H_0 \to \tau \bar \mu)$ versus
$M_0=M_{1/2}(GeV)$ for $\theta_2 =\pi e^{0.8i}$. 
{\bf (4f)} Same as (4e) but for $BR(\tau \to \mu \gamma)$.  
The horizontal lines are the upper experimental bound on $BR(\tau \to \mu \gamma)$. 
\label{fig:4}}
\end{figure}

We show in figs.~(\ref{fig:1}) through ~(\ref{fig:4}) the numerical results for the branching ratios of
the LFVHD together with the branching ratios for the relevant $l_j \to l_i \gamma$ decays.
The results of $BR(H_0 \to \tau \bar \mu)$ 
as a function of $m_N$, for degenerate heavy neutrinos and real R, are illustrated in 
fig.~(\ref{fig:1}a), for several $\tan\beta$ values, $\tan\beta=3,10,30,50$. Notice that in
this case, the rates do not depend on $R$. 
The explored range in $m_N$ is
from $10^8 $ GeV up to $10^{14} $ GeV which is favorable  
for baryogenesis. We also show in this figure, the
corresponding 
predicted rates for the most relevant lepton decay,
 which in this case is $\mu \to e \gamma$, and include its upper
 experimental bound. We have checked that the other lepton decay channels are well within
 their experimental allowed range.  The ratios for $A_0$ decays, not shown here for brevity, 
  are very similar to those for 
 $H_0$ decays in all the studied scenarios in this work. We have also found that the  
ratios for the light Higgs boson, $h_0$, behave very similarly with $m_N$ and $\tan\beta$ but 
are smaller than the 
heavy Higgs ones in  about two orders of magnitude. 
>From our results  we learn about the  high sensitivity 
to $\tan\beta$ of the LFVHD rates for all Higgs
bosons which, at large $\tan \beta$, scale roughly as $(\tan\beta)^4$, in comparison
with the lepton decay rates which scale as $(\tan\beta)^2$. The dependence of both rates
on $m_N$ is that expected from the mass insertion approximation, where 
$BR(H_{x}\to l_j {\bar l_i})$, $BR (l_j \to l_i \gamma) \propto |m_N\log(m_N)|^2 $.
We find that the largest ratios, which are for $H_0$ and $A_0$, are in any case 
very small, at most $10^{-10}$ in the region of high $\tan \beta$ and high $m_N$. 
Besides, the rates for $\mu \to e \gamma$ decays are below the upper
experimental bound for all explored $\tan\beta$ and $m_N$ values.    
The branching ratios for the Higgs boson decays into $\tau\bar{e}$ and $\mu \bar e$ 
are much smaller
than the $\tau\bar{\mu}$ ones, as expected,  and we do not show plots for them. 
For instance, for 
$m_N=10^{14}$ GeV,
and $\tan\beta=50$ we find $BR(H^{(x)} \to \tau {\bar\mu})/BR(H^{(x)} 
\to \tau {\bar e})=4 \times 10^3$ and 
$BR(H^{(x)} \to \tau {\bar\mu})/BR(H^{(x)} \to \mu {\bar e})=1.2 \times 10^6$ for 
the three Higgs bosons. 

The case of hierarchical neutrinos gives clearly larger LFV rates than the degenerate case, as
can be seen in figs.~(\ref{fig:2}),~(\ref{fig:3}) and~(\ref{fig:4}). However, we will get 
restrictions on 
the maximum allowed Higgs decay rates coming from the experimental lepton 
decay bounds. 
For instance, the case of real $\theta_1$, that is illustrated in fig.~(\ref{fig:2}) 
shows that compatibility with $\mu \to e \gamma$ data occurs
 only in the very narrow deeps at  around
 $\theta_1=0$, $1.9$ and $\pi$. Notice that 
it is precisely at the points $\theta_1=0, \pi$ where the  
$BR (H_0, A_0 \to \tau \bar \mu)$ rates
reach their maximum values, although these are not large, just about $10^{-8}$. 
We have checked that for lower $\tan\beta$ values, the allowed regions in $\theta_1$ widen 
and are placed at the same points, but the corresponding maximum values of the LFVHD rates 
get considerably reduced. For
the alternative case, not shown here,  of real $\theta_2 \neq 0$, with $\theta_1=\theta_3=0$ we
get a similar behaviour of   
$BR (H_x \to \tau \bar \mu)$ with $\theta_2$ than with  $\theta_1$, and the maximum values 
of about $10^{-8}$ are now placed at 
$\theta_2=0$, $\pi$. In contrast, $BR(\mu \to e \gamma)$ is constant with 
$\theta_2$ and reach very small values, well below the experimental bound. 
In particular, for $\tan \beta =50$, $M_0=400$ GeV and
$M_{1/2}=300$ GeV it is $10^{-19}$. 
Regarding the dependence with $\theta_3$, not shown here either, a reverse situation 
is found, where  $BR (H_x \to \tau \bar \mu)$ is approximately constant and, for the 
heavy Higgs bosons, it is around $10^{-8}$. On the contrary, 
$BR(\mu \to e \gamma)$ varies but it is always well below the experimental upper bound. 
In addition, we have checked that the $BR(\tau \to \mu \gamma)$ and $BR(\tau \to e \gamma)$  rates are
within the experimental allowed range in all cases. 
In conclusion, for real R we find that the maximum allowed LFVHD rates are 
at or below $10^{-8}$. 
The case of complex $R$ is certainly more promissing. The examples illustrated in 
figs.~(\ref{fig:3}) and~(\ref{fig:4}) are for the most favorable case, among the ones
studied here, of complex $\theta_2 \neq 0$ with $\theta_1=\theta_3=0$ and show that 
considerably larger $BR (H_x \to \tau \bar \mu)$ rates than in the real $R$ case 
are found. Regarding the dependence with $\theta_2$, we find that for the explored  values
with  $(|\theta_2|, Arg(\theta_2)) \leq (3.5,1)$,
the Higgs rates 
grow with both $|\theta_2|$ and $Arg (\theta_2)$ and, for the selected values of the
MSSM-seesaw
parameters 
in fig.~(\ref{fig:3}), they reach values up to 
around $5 \times 10^{-5}$. We have checked that the predicted rates for $BR(\mu \to e
\gamma)$  are well below the experimental upper bound, being nearly
constant with $\theta_2$ and around $10^{-19}$. Similarly, for the $\tau \to e \gamma$
decay. Notice that the smallness of these two decays, in the case under study of $\theta_2 \neq 0$, 
is not maintained if our hypothesis on $\theta_{13}=0$ is changed. For instance, for 
$\theta_{13}=5^o$, which is also allowed by neutrino data, 
we get $BR(\mu \to e \gamma)\sim 1.8 \times 10^{-8}$ well above the
experimental upper bound.
Therefore, in this case of complex $\theta_2 \neq 0$, the relevant 
lepton decay is $\tau \to \mu \gamma$ which
is illustrated in figs.~(\ref{fig:3}) and ~(\ref{fig:4}) together with its experimental
bound. For the set of parameters chosen in fig.~(\ref{fig:3}), we get 
that the allowed region by $\tau \to \mu \gamma$ data of the 
$(|\theta_2|, Arg(\theta_2))$
parameter space implies a reduction in the Higgs rates, leading to a maximum 
allowed value of just $5 \times 10^{-8}$. 
The dependence of the LFV ratios with $M_0$ and $M_{1/2}$ for hierarchical neutrinos are 
shown in fig.~(\ref{fig:4}). We see clearly the different behaviour of the LFVHD and the lepton
decays with these parameters, showing the first ones a milder dependence.  
This implies, that for large enough values of $M_0$ or $M_{1/2}$ or both the 
$BR(\tau \to \mu \gamma)$ rates get considerably suppresed, due to the decoupling of 
the heavy SUSY
particles in the loops, and enter into the allowed region by data, whereas the   
$BR(H_0\to\tau {\bar \mu})$ rates are not much reduced. In fact, we see in 
figs.~(\ref{fig:4}e) and ~(\ref{fig:4}f) that for the choice $M_0=M_{1/2}$ the $\tau$ 
decay ratio
crosses down the upper experimental bound at around $M_0=1100$ GeV whereas the 
Higgs decay ratio is
still quite large $\sim 4 \times 10^{-6}$ in the high $M_0$ region, around 
$M_0 \simeq 2000$ GeV.  
This behaviour  is a clear indication that the heavy SUSY particles in the loops do not
decouple in the LFVHD.
Notice also that it can be reformulated 
as non-decoupling in the effective $H^{(x)} \tau \mu$ couplings and these in turn can 
induce large contributions to other LFV processes that are mediated by Higgs exchange as, 
for instance, $\tau \to \mu \mu \mu$. However, we have checked that for 
the explored values in this work of $M_0$, $M_{1/2}$, $\tan \beta$, R and $m_{N_i}$ that 
lead to the anounced LFVHD ratios of about $4 \times 10^{-6}$, the corresponding 
$BR(\tau \to \mu \mu \mu)$ rates are below the present experimental upper 
bound. 

In summary, after exploring the dependence of the LFVHD rates with all the involved MSSM-seesaw
parameters, 
and by requiring 
compatibility with data of the correlated predictions for $\mu \to e \gamma$, $\tau \to e \gamma$
and $\tau \to \mu \gamma$  decays, we find that  $BR(H_0, A_0 \to \tau {\bar \mu})$  as large as $~10^{-5}$, for hierarchical neutrinos and large $M_{SUSY}$ in the TeV range can be reached. 
These rates are
close but still below the expected future experimental reach of about $10^{-4}$ at the 
LHC and next
generation linear colliders. 
\section*{Acknowledgments}
M.J. Herrero wishes to thank the organizers of this conference for a very fruitful,
interesting and enjoyable meeting.

\end{document}